\documentclass[12pt,english]{article}
\usepackage[a4paper,bindingoffset=0.2in,%
left=1in,right=1in,top=1in,bottom=1in,%
footskip=.25in]{geometry}
\usepackage[super,compress]{cite}
\usepackage{hyperref}
\hypersetup{colorlinks,urlcolor=black,citecolor=black,linkcolor=black,filecolor=black}
\usepackage{graphicx}	% Including figure files
\usepackage{amsmath}	% Advanced maths commands
\usepackage{amssymb}	% Extra maths symbols
 \usepackage{breakurl}
\newcommand{\be}{\begin{equation}}
\newcommand{\ee}{\end{equation}}
\newcommand{\ds}{\displaystyle}

\begin{document}
	
	\markboth{Rohin Kumar}
	{A New Class of Cosmologically `Viable' $f(R)$ Models}
	
	%\catchline{}{}{}{}{}
	%
	%%%%%%%%%%%%%%%%%%%%%%%%%%%%%%%%%%%%%%%%%%%%%%%%%%%%%%%%%%%%%%%%%%%%
	
	\title{A New Class of Cosmologically `Viable' $f(R)$ Models}
	
	\author{Rohin Kumar\footnote{Department of Physics \& Astrophysics, 
			University of Delhi, North Campus, 
			Delhi, India}}
	
%	\address{University of Delhi, North Campus\\
%		Delhi, India\\
%		yrohinkumar@gmail.com}
	
%	\maketitle
	
%	\begin{history}
%		\received{$29^{th}$ September 2016}
%		\revised{$16^{th}$ November 2016}
%	\end{history}

%	\keywords{Linear Coasting Cosmology; Designer $f(R)$ gravity; new class of viable $f(R)$.}
%	\keywords{Linear Coasting; Designer $f(R)$; $f(R)$ gravity; alternative cosmology.}
%	\ccode{PACS numbers: 04.50.Kd, 98.80.Jk}

\maketitle
	\begin{abstract}
	Instead of assuming a form of gravity and demand cosmology fit with $\Lambda CDM$, a potentially `viable' $f(R)$ gravity model is derived assuming an alternative model of cosmology. Taking the `designer' approach to $f(R)$, a new class of solutions are derived starting with linear coasting cosmology in which scale factor linearly increases with time during matter domination. The derived forms of $f(R)$ are presented as result.
\end{abstract}
\section{Introduction}
\label{sec:intro}

Alternative gravity models have been proposed numerous times to resolve the problems of dark matter, dark energy and to address other issues of Cosmology such as inflation. $f(R)$ theories have been of much interest as toy-models in exploring alternative gravity cosmologies\cite{sotiriou2010f}. Some,\cite{napolitano2012testing,Stabile:2013jon} through $f(R)$ models, even made attempts to explain other effects of dark matter such as flat rotation curves of galaxies without needing any exotic matter.  

In context of cosmology, $f(R)$ models are generally studied as possible alternatives to either dark energy or dark matter. There are some $f(R)$ models that mimic $\Lambda CDM$ behavior under certain limits.  Their viability is mostly judged on a model's ability to reproduce scale factor evolution as predicted by $\Lambda$CDM model. However, there seems to be no one definitive $f(R)$ model that possibly satisfies all the required criteria to be a strong contender to $\Lambda$CDM model\cite{sotiriou2010f}. 
 
The aim of this paper is to explore the possibility of new viable $f(R)$ models assuming a universe with scale factor linearly evolving with time (at least during matter domination). This paper starts with brief background introduction to $f(R)$ gravity. After presenting generic field equations of $f(R)$ gravity we write modified FRW equations for any $f(R)$ model to study the cosmology of an isotropic and homogeneous universe. The idea of `designer' approach to $f(R)$ gravity and linear coasting model are introduced in the subsequent sections. Using the designer approach, functional forms of $f$ for different geometries of a linearly coasting Universe are derived in the final section. Discussion of the results is followed by conclusions and plan for future work.

\section{$f(R)$ Cosmology}

Just as one can derive the field equations of general relativity from Einstein-Hilbert action, one can assume any functional form $f$ in terms of Ricci scalar($R$) and it could as well have been an alternative model of gravity. Although adding quadractic terms etc. to the Lagrangian was done just after few years of introduction of GR, the motivation to modify gravity has become more compelling in the recent times in light of latest cosmological observations.

In the simplest form of $f(R)$, one can simply extremize the action with respect to the metric tensor alone. This is called the metric formalism of $f(R)$ theory. We write the action as 
\be 
S={1 \over 2\chi} \int d^4x\sqrt{-g}f(R)+S_{matter}
\ee
We extremize this action w.r.t $\delta g^{\alpha\beta}$ to write $\delta S=0$. After applying relevant boundary conditions, we get the generic field equations of the form
\be
f'(R)R_{\mu\nu}-
\frac{1}{2}f(R)g_{\mu\nu}
-(\nabla_\mu\nabla_\nu-g_{\mu\nu}\square)f'(R)=\chi T_{\mu\nu}
\label{gfe}
\ee  
Here $\chi=8\pi G$ where `G' is universal gravitational constant. $g$ is the determinant of the metric and $R$ is the Ricci scalar. Working in natural units, we, have $c=\hbar=1$. 

The trace of equation \eqref{gfe} gives us 
\be
3\square f'(R)+Rf'(R)-2f(R)=\chi T
\label{trace}
\ee
Here $T$ is the trace of the matter stress-energy tensor $T_{\mu\nu}$. %This equation gives us insights into some basic properties of $f(R)$ theories in some cases.
 
Starting with homogeneous and isotropic Universe, we take the FLRW metric element where $a(t)$ is the scale factor of the Universe.
\be 
ds^2=-dt^2+a^2(t)\left(\dfrac{dr^2}{1-\kappa r^2}+r^2(d\theta^2+\sin^2\theta d\phi^2)\right)
\ee

for a generic $f(R)$ model, we get modified Friedmann equations as
 
\be 
H^2+\dfrac{\kappa}{a^2}=\dfrac{1}{3f'(R)}\left[\chi \rho+\dfrac{Rf'(R)-f(R)}{2}-3Hf''(R)\dot{R}\right]
\label{modfrw1}
\ee

and 

\begin{multline} 
2\dot{H}+3H^2+\dfrac{\kappa}{a^2}=-\dfrac{1}{f'(R)}\bigg[\chi P+2H\dot{R}f''(R)+\\
\dfrac{f(R)-Rf'(R)}{2}+\ddot{R}f''(R)+\dot{R}^2f'''(R)\bigg]\label{modfrw2}
\end{multline}

	The effective density and pressure profiles due to $f(R)$ can be used to write the FRW equations in their standard cosmological form.
	\be 
	H^2+\dfrac{\kappa}{a^2}=\dfrac{1}{3f'(R)}(\chi\rho+\rho_{eff})\label{modfrw1_eff}
	\ee
	
	\be 
	2\dot{H}+3H^2+\dfrac{\kappa}{a^2}=-\dfrac{1}{f'(R)}(\chi P+P_{eff})
	\ee
	Here $\rho_{eff}$ and $P_{eff}$ are additional contributions to density and pressure profiles because of $f(R)$
	\be 
	\rho_{eff}=\dfrac{Rf'(R)-f(R)}{2}-3Hf''(R)\dot{R}
	\ee
	
	\be 
	P_{eff}=2H\dot{R}f''(R)+\dfrac{f(R)-Rf'(R)}{2}+\ddot{R}f''(R)+\dot{R}^2f'''(R)
	\ee

As one can quickly note from \eqref{modfrw1} and \eqref{modfrw1_eff}, one cannot write the total of energy densities in fraction of critical density as `$1$' as we do in standard cosmology\cite{weinberg2008cosmology}. Although, some dynamical system of equations can be written\cite{sotiriou2010f} they are not readily solvable.

\be 
w_{eff}=\dfrac{P_{eff}}{\rho_{eff}}=\dfrac{\dot{R}^2f'''+2H\dot{R}f''+\ddot{R}f''+\frac{1}{2}(f-Rf')}{\frac{Rf'-f}{2}-3H\dot{R}f''}
\label{weff}
\ee

The effective pressure and density terms because of $f(R)$ are likened to contributions of unknown parameters like dark energy in the standard model equations. To judge how well a model fits with the observations, generally some limit to $\Lambda$CDM model is sought. Many attempts presented in reviews \cite{de2010f,sotiriou2010f,fR_thesis} take a similar approach to liken effective $w$ term ($w_{eff}$) to dark energy in various versions of $f(R)$ eliminating them based on the fitness with dynamics of $\Lambda$CDM model.
 
\section{The `Designer' Approach to $f(R)$}
In the standard approach, some form of $f(R)$ is assumed and a fit with $\Lambda CDM$ is expected/produced as a consequence. Such theories, try to achieve GR limit by reproducing standard cosmology with $f(R)$ as the dark energy or early inflation replacement. In general, any model of $f(R)$ looking to explain late-time acceleration is expected to give rise to a cosmology which also preserves the evolution sequence of the standard model viz. early inflation, radiation domination era (during which Big Bang Nucleosynthesis occurs), a matter dominated era and the present accelerated epoch.

However, one could, in principle start with the scale factor $a(t)$ and look to find the suitable functional forms of $f(R)$. One can prescribe the desired form of the scale factor $a(t)$ and integrate a differential equation for $f(R)$ that produces the desired scale factor. These are  the so-called `Designer' $f(R)$ gravity models. This approach of reconstruction of $f(R)$ from expansion is pioneered and explored further by Nojiri and Odintsov\cite{nojiri2006modified}\cite{nojiri2007modified}\cite{nojiri2009cosmological} . For $\Lambda CDM$, we have to resort to numerical techniques to look for forms of $f$\cite{song2007large}. These models, often need fine-tuning of constants\cite{pogosian2008pattern}. Dunsby \textit{et al.}\cite{dunsby2010lambda}, after studying various reconstructions of $f(R)$ gravity for FRW expansion history, have concluded that only simple function of Ricci scalar $R$ that admits an exact $\Lambda CDM$ expansion history is standard General Relativity with cosmological constant and additional degrees of freedom added to matter term. Moreover, the prescribed evolution of the scale factor $a(t)$ does not uniquely determine the form of $f(R)$ but could possibly give rise to a class of $f(R)$ models that need to be further explored and constrained from observational data\cite{sotiriou2010f}. 

\section{Linear Coasting Cosmology}

 When we look beyond the standard model fit to cosmology, we could possibly find some interesting forms of $f(R)$ using the designer approach. One such model, which has a decent observational fit to cosmology is a universe with linearly coasting scale factor. Linear coasting cosmology is one of the many attempts that tried to end the `Dark' age of fundamental cosmology with a coherent theoretical construct. Even recent statistical analysis of the Supernovae Ia data\cite{nielsen2015marginal} support such simple yet effective model of the Universe. Hence, linear coasting model fits Supernovae data as good as the $\Lambda$CDM model without commonly expected late-time acceleration. This model of the Universe was studied extensively by Daksh \textit{et al.} in Refs. \citen{dlohiya2000,dev2001linear,dev2002cosmological,gehlaut2002freely}.
 
 Apart from solving the cosmological constant problem and presenting a simmering big bang nucleosynthesis, this model offers solutions to the age problem and horizon problem\cite{gehlaut2002freely}. As conventionally expected, Milne model (i.e. an empty open Universe) is not the only one with linearly coasting scale factor. One could achieve linear coasting by modified gravity also as suggested by Daksh \textit{et al.}\cite{dlohiya2000} as one of the possible motivations to their work, although primarily $a(t)\propto t$ is taken as an ansatz. This paper intends to derive some gravity models that could support the linear coasting model. Recently, there is lot of interest in a model on similar lines as linear coasting model called $R_h=ct$ model which also has a linearly evolving scale factor with $\ddot{a}=0$ with $w=-1/3$. For more details please see Melia \textit{et al.} in Refs. \citen{melia2012fitting,melia2012rh,wei2015comparative}. While comparisons of the results with this model might be inevitable, comments and comparisons will be subject matter of another paper.

\section{$f(R)$ of a Linearly Coasting Universe}

For a linearly coasting scale factor we have 
\be
a(t)\propto t \implies a(t)=t/t_0
\ee

This gives us 
\be
H(t)={1 \over t}~~~\&~~~H_0={1\over t_0}
\ee

For FLRW metric Ricci scalar is
\be
R=6(2H^2+\dot{H}+\kappa/a^2) 
\ee

%\Big{}
%\Large{(\textbf{i})~$k=0$}

\subsection{For $\kappa=0$:}

Assuming a flat Universe, we have 

\be
R=6(2H^2+\dot{H}) = {6 \over t^2} = 6H^2 \implies H=\sqrt{R/6}\label{Rk0}
\ee

Using this expression we can convert the modified Friedmann equations \eqref{modfrw1} and \eqref{modfrw2} as functions of $R$.

This leads us to 
\be
R^2f''-f/2+\chi\rho=0\label{fR1k0}
\ee
\be
2R^3f'''+R^2f''-Rf'+{3 \over 2}f+3\chi P =0\label{fR2k0}
\ee
We do not get any new information from the trace equation as it is same as \eqref{fR1k0}. 

Since we are primarily interested in late-time observational data (such as supernovae data), we can assume the case of matter dominated era where $P=0$. Simplifying these two equations, we get a $2^{nd}$ order differential equation

% making $w=0$ in \eqref{fR1}

%assuming the equation of state $w=P/\rho$
\be
f''-{1\over 2R^2}f+\alpha R^{-1/2}=0\label{difffRk0}
\ee

Here we use the fact that (also using \eqref{Rk0})

%\begin{eqnarray}
$$\rho = {\rho_0 a_0^3 \over a^3} = {\rho_0 t_0^3 \over t^3} $$
$$\chi\rho = \alpha R^{3/2}$$
%\end{eqnarray}

%we write in \eqref{difffRk0}

where $\alpha = \ds{\chi\rho_0 t_0^3 \over 6\sqrt{6}}$ or 

$$\alpha = {\ds\chi\rho_0 \over 6\sqrt{6}H_0^3}={4\pi G\rho_0 \over 3\sqrt{6}H_0^3}$$

Solving \eqref{difffRk0}, we get the general form of $f(R)$ as

%Solving \eqref{difffRk0} we get a possibly viable form for $f(R)$
%
%\be
%f(R)=-{4\alpha \over 1 + 3 w (4 + 3 w)} R^{{3 \over 2}(1+w)}+C_1R^{(\sqrt{3}+1)/2}+C_2R^{(-\sqrt{3}+1)/2}\label{fR1}
%\ee

% giving us

\be
f(R)=-4\alpha R^{3/2}+C_1R^{(\sqrt{3}+1)/2}+C_2R^{(-\sqrt{3}+1)/2}\label{fR2}
\ee

Despite its contrived appearance, this is potentially a viable form of $f(R)$ with constants $C_1$, $C_2$ along with $\rho_0$(and hence $\alpha$) that need to be constrained using observational data. 

Fearing divergences at $R\rightarrow 0$ we can set $C_2=0$ resulting in

\be
f(R)=C_1R^{(\sqrt{3}+1)/2}-4\alpha R^{3/2}\label{fR2_}
\ee

%For the sake of completion, if the Universe was primarily radiation dominated ($w=1/3$)at late times we get 
%
%\be
%f(R)=-2/3\alpha R^2+C_1R^{(\sqrt{3}+1)/2}+C_2R^{(-\sqrt{3}+1)/2}\label{fR1}
%\ee

\subsection{For $\kappa>0$:}

In case of the closed universe, we have $\kappa/a^2 = 1/t^2$ for linear coasting. This gives us
\be
R=6(2H^2+\dot{H}+\kappa/a^2) = 12/t^2 = 12H^2 \implies H =\sqrt{R/12}
\ee
Rest of the equations follow this definiton of Ricci scalar giving rise to
\be
R^2f''+Rf'-f+2\chi\rho=0
\ee
\be
R^3f'''+{R^2 \over 2}f''-{3 \over 2}Rf'+{3 \over 2}f+3\chi P = 0
\ee

Solving these two equations, we have a simpler $1^{st}$ order differential equation unlike \eqref{difffRk0} as
\be
f'-{1 \over R}f+{\beta \over 2}R^2=0\label{difffRk1}
\ee

Here

$$\chi\rho = \beta R^{3/2}$$

where $\beta = \ds{\chi\rho_0 t_0^3 \over 24\sqrt{3}}$ or 

$$\beta = {\ds\chi\rho_0 \over 24\sqrt{3}H_0^3}={\pi G\rho_0 \over 3\sqrt{3}H_0^3}$$

%Solving this we get 
%\be
%f(R)=C_1R-{2\alpha \over (1 + 3 w)} R^{{3 \over 2}(1 + w)}
%\ee

Solving \eqref{difffRk1}, we get solution for $f(R)$ as

\be
f(R)=C_1R-2\beta R^{3/2}\label{fR3_}
\ee

%In case of radiation dominated Universe, we have
%\be
%f(R)=C_1R-\alpha R^2
%\ee

\subsection{For $\kappa < 0$:}

Unfortunately in the open Universe case, $\kappa/a^2 = -1/t^2$ gives 

\be
R=6(2H^2+\dot{H}+\kappa/a^2) = 0
\ee

Both L.H.S and R.H.S of modified Friedmann equations \eqref{modfrw1} \eqref{modfrw2} will be $0$ as all are written as functions of $R$. One can, in principle, assume a power-law solution of scale factor $a(t)\propto t^m$ in which $m\approx 1$. Then taking 
$$H(t)=\dfrac{\dot{a}}{a}=\dfrac{mt^{m-1}}{t^m}=\dfrac{m}{t}$$ 
one can look for possible forms of $f(R)$. In fact, such an analysis can be done for the flat and closed cases also. Unfortunately, this leads us to a set of transcendental equations that are not readily solvable.
%(not even numerically) without fixing the arbitrary constants.

As for the solutions of $f(R)$ found, we only consider cosmological viability with linear coasting as a possible alternative model of cosmology. In general, any $f(R)$ model is considered `viable' if it fulfills the following criteria as given in Ref. \cite{faraoni2014nine}.
\begin{enumerate}
	\item produces correct cosmological dynamics
	\item is stable without ghosts in theory
	\item has well-posed Cauchy problem
	\item gives correct weak-field limit (Newtonian \& post-Newtonian)
	\item has cosmological perturbation theory compatible with CMB and large-scale observations
	%As a preliminary test of viability, criteria (i)-(iii) are looked in this paper.  
\end{enumerate}
 The Cauchy problem for general metric $f(R)$ is ascertained to be well-posed\cite{capozziello2012cauchy}. The other viability conditions such as stability (needing $f''(R)\geq 0$ etc.) and weak-field limit are studied for models that are close to GR limit\cite{faraoni2014nine}. This makes solution given in \eqref{fR3_} unstable if $C_1=1$. But stability of these solutions in \eqref{fR2_} and \eqref{fR3_} needs to be studied as they are not readily in $R+\epsilon \phi(R)$ form\cite{faraoni2014nine}. While some of the results may be familar forms of $f(R)$ having fractional powers of $R$\cite{broy2015disentangling}\cite{dunsby2010lambda}, they still need to be explored in the context of linear coasting model. Also, one needs to analyze the weak-field limit and stability of the linear perturbation solutions\cite{sawicki2007stability} of the derived $f(R)$ forms to consider them `viable'.
 % and check under what conditions the theory approches general relativity (GR) limit. 
 
 With well-defined alternative $f(R)$ models of gravity in place, one should be able to carry out calculations to further investigate the structure formation and CMB spectrum results to find if these models are strong contenders for an alternative model of cosmology. One could, in principle, substitute these $f(R)$ forms in \eqref{modfrw1} and \eqref{modfrw2} to numerically obtain evolution of scale factor $a(t)$ during radiation domination given proper initial conditions. As late time/matter dominated solutions of $f(R)$ are not same as the radiation dominated cases, one needs to see how the radiation domination era evolves and if it produces relevant predictions in terms of nucleosynthesis and other early universe expected behavior. However, if linear coasting goes back to early enough epochs, simmering nucleosynthesis can still take place\cite{gehlaut2002freely}.
%Some of the results derived above may be familar forms of $f(R)$, but they still need to be explored in the context of linear coasting model.
\section{Conclusions \& Future Work}
Assuming a linearly coasting scale factor, we derived a potentially new `viable' forms of $f(R)$. While these forms may not look suitable in terms of $\Lambda CDM$ or conventional $f(R)$ theory, they need to be re-evaluated in the light of linear coasting cosmology. As there are fewer options on constraining $f(R)$ models other than Cosmology, one can look at linear growth rate of structures and gravitational weak lensing\cite{tsujikawa2009dispersion} as possible  observations to constrain these new classes of models. There is also need for scrutinizing stability of such solutions. CMB and structure formation theories for these models need to be studied. One can also evaluate the efficacy of these new forms in the weak field limit from the solar system tests and gravitational wave observations. These areas are to be explored in the subsequent work(s).
%\acknowledgments
\section*{Acknowledgments}
I thank my supervisors Prof. Daksh Lohiya \& Prof. Amitabha Mukherjee for giving me freedom to pursue my own ideas along with necessary feedback. I thank CSIR  (sanction no. 09/045(1324)/2014-EMR-I) for providing financial support for my research work. Special thanks are due to faculty involved in DST, India grant project SR/S2/HEP-12/2006 for encouraging my work with (partial) financial support when I had none.
%supporting my research efforts

\bibliographystyle{plain}
\bibliography{1611.03728v3.bib}

\begin{thebibliography}{10}

\bibitem{broy2015disentangling}
Benedict~J Broy, Francisco~G Pedro, and Alexander Westphal.
\newblock Disentangling the f (r)-duality.
\newblock {\em Journal of Cosmology and Astroparticle Physics}, 2015(03):029,
  2015.

\bibitem{capozziello2012cauchy}
S~Capozziello and S~Vignolo.
\newblock The cauchy problem for {$f(R)$}-gravity: An overview.
\newblock {\em International Journal of Geometric Methods in Modern Physics},
  9(01):1250006, 2012.

\bibitem{de2010f}
Antonio De~Felice and Shinji Tsujikawa.
\newblock ${f(R)}$ theories.
\newblock {\em Living Rev. Rel}, 13(3):1002--4928, 2010.

\bibitem{dev2002cosmological}
Abha Dev, Margarita Safonova, Deepak Jain, and Daksh Lohiya.
\newblock Cosmological tests for a linear coasting cosmology.
\newblock {\em Physics Letters B}, 548(1):12--18, 2002.

\bibitem{dev2001linear}
Abha Dev, Meetu Sethi, and Daksh Lohiya.
\newblock Linear coasting in cosmology and {SNe Ia}.
\newblock {\em Physics Letters B}, 504(3):207--212, 2001.

\bibitem{dlohiya2000}
M.~Sethi D.Lohiya, A.~Batra.
\newblock Comment on ``observational constraints on power-law cosmologies".
\newblock {\em Phys. Rev. D}, 60(10):108301, 1999.

\bibitem{dunsby2010lambda}
Peter~KS Dunsby, Emilio Elizalde, Rituparno Goswami, Sergei Odintsov, and Diego
  Saez-Gomez.
\newblock $\lambda$ cdm universe in f (r) gravity.
\newblock {\em Physical Review D}, 82(2):023519, 2010.

\bibitem{faraoni2014nine}
Valerio Faraoni.
\newblock Nine years of {$f(R)$} gravity and cosmology.
\newblock In {\em Accelerated Cosmic Expansion}, pages 19--32. Springer, 2014.

\bibitem{gehlaut2002freely}
Savita Gehlaut, A~Mukherjee, S~Mahajan, and D~Lohiya.
\newblock A ``freely coasting" universe.
\newblock {\em arXiv preprint astro-ph/0209209}, 2002.

\bibitem{fR_thesis}
Alejandro Guarnizo.
\newblock Cosmological models in modified ${f(R)}$ gravity theories.
\newblock {\em arXiv preprint arXiv:1211.2444}, 2012.

\bibitem{melia2012fitting}
Fulvio Melia.
\newblock Fitting the {Union2.1 Supernova Sample} with the ${R_h= ct}$
  universe.
\newblock {\em The Astronomical Journal}, 144(4):110, 2012.

\bibitem{melia2012rh}
Fulvio Melia and ASH Shevchuk.
\newblock The ${R_h= ct}$ universe.
\newblock {\em Monthly Notices of the Royal Astronomical Society},
  419(3):2579--2586, 2012.

\bibitem{napolitano2012testing}
NR~Napolitano, S~Capozziello, AJ~Romanowsky, M~Capaccioli, and C~Tortora.
\newblock Testing yukawa-like potentials from ${f(R)}$-gravity in elliptical
  galaxies.
\newblock {\em The Astrophysical Journal}, 748(2):87, 2012.

\bibitem{nielsen2015marginal}
Jeppe~Tr{\o}st Nielsen, Alberto Guffanti, and Subir Sarkar.
\newblock Marginal evidence for cosmic acceleration from {Type Ia} supernovae.
\newblock {\em arXiv preprint arXiv:1506.01354}, 2015.

\bibitem{nojiri2007modified}
Shin'ichi Nojiri and Sergei~D Odintsov.
\newblock Modified gravity and its reconstruction from the universe expansion
  history.
\newblock In {\em Journal of Physics: Conference Series}, volume~66, page
  012005. IOP Publishing, 2007.

\bibitem{nojiri2009cosmological}
Shin'ichi Nojiri, Sergei~D Odintsov, and Diego S{\'a}ez-G{\'o}mez.
\newblock Cosmological reconstruction of realistic modified f (r) gravities.
\newblock {\em Physics Letters B}, 681(1):74--80, 2009.

\bibitem{nojiri2006modified}
Shin’ichi Nojiri and Sergei~D Odintsov.
\newblock Modified f (r) gravity consistent with realistic cosmology: From a
  matter dominated epoch to a dark energy universe.
\newblock {\em Physical Review D}, 74(8):086005, 2006.

\bibitem{pogosian2008pattern}
Levon Pogosian and Alessandra Silvestri.
\newblock Pattern of growth in viable {$f(R)$} cosmologies.
\newblock {\em Physical Review D}, 77(2):023503, 2008.

\bibitem{sawicki2007stability}
Ignacy Sawicki and Wayne Hu.
\newblock Stability of cosmological solutions in {$f(R)$} models of gravity.
\newblock {\em Physical Review D}, 75(12):127502, 2007.

\bibitem{song2007large}
Yong-Seon Song, Wayne Hu, and Ignacy Sawicki.
\newblock Large scale structure of {$f(R)$} gravity.
\newblock {\em Physical Review D}, 75(4):044004, 2007.

\bibitem{sotiriou2010f}
Thomas~P Sotiriou and Valerio Faraoni.
\newblock ${f(R)}$ theories of gravity.
\newblock {\em Reviews of Modern Physics}, 82(1):451, 2010.

\bibitem{Stabile:2013jon}
A.~Stabile and S.~Capozziello.
\newblock Galaxy rotation curves in ${f(R,\phi)}$ gravity.
\newblock {\em Phys.Rev.}, D87(6):064002, 2013.

\bibitem{tsujikawa2009dispersion}
Shinji Tsujikawa, Radouane Gannouji, Bruno Moraes, and David Polarski.
\newblock Dispersion of growth of matter perturbations in ${f(R)}$ gravity.
\newblock {\em Physical Review D}, 80(8):084044, 2009.

\bibitem{wei2015comparative}
Jun-Jie Wei, Xue-Feng Wu, Fulvio Melia, and Robert~S Maier.
\newblock A comparative analysis of the supernova legacy survey sample with
  ${\Lambda}${CDM} and the ${R_h= ct}$ universe.
\newblock {\em The Astronomical Journal}, 149(3):102, 2015.

\bibitem{weinberg2008cosmology}
Steven Weinberg.
\newblock {\em Cosmology}.
\newblock Oxford University Press, 2008.

\end{thebibliography}
\end{document}